\documentclass[preprint,aps,prl,floats]{revtex4}%
\usepackage{epsfig}
\usepackage{dcolumn}
\usepackage{bm}
\usepackage{amsmath}
\usepackage{amsfonts}
\usepackage{amssymb}
\usepackage{graphicx}
\newcommand{\beq}{\begin{eqnarray}}
\newcommand{\eeq}{\end{eqnarray}}

\begin{document}
\title{$U-J$ Synergy Effect for the High  $T_c$ Superconductors}
\author{ Liliana Arrachea$^{1,2}$ and Dra\v{z}en Zanchi$^{2}$. }
\affiliation{$^{1}$  Departamento de F\'{\i}sica, Universidad de
Buenos Aires, Ciudad Universitaria Pabell\'on I, (1428) Buenos Aires,
Argentina.\\
$^{2}$
 Laboratoire de Physique Th{\'e}orique  et Hautes {\'E}nergies,  
4 Place Jussieu, 75252 Paris Cedex 05, France}

\begin{abstract}
Using renormalization group with included self-energy effects
and exact diagonalization of small clusters
we investigate the ground state phase diagram
of a two-dimensional extended Hubbard model with
nearest-neighbor exchange interaction $J$, in addition to the local Coulomb
repulsion $U$.
The main instabilities are antiferromagnetism
close to half-filling and
$d_{x^2-y^2}$ superconductivity in the doped system.
We find that self-energy effects are fatal for superconductivity in the
repulsive Hubbard model (i.e. $J=0$, $U>0$). The superconductivity is
{\em triggered} by finite $J$.
The combined action of $J$ and $U$ interactions provide a
remarkably efficient mechanism to
enhance both $d_{x^2-y^2}$ superconducting and antiferromagnetic
correlations.
\end{abstract}


\maketitle

\section{Introduction}
One of the most striking features observed in the
phase diagram of the high T$_c$ superconducting cuprates is the proximity between the
insulating state with long-range antiferromagnetic order and the superconducting 
phase.  This remarkable issue and the fact that at optimum doping the 
antiferromagnetic coherence length remains finite ($\xi \sim 4a$ for 
La$_{2-x}$Sr$_x$Cu0$_4$),  
have been the source of inspiration for many
theoretical works in which the pairing mechanism is proposed to be
originated in the fluctuations of the spin-density wave phase. 
Another peculiarity of these materials is that the superconducting
order parameter has $d_{x^2-y^2}$ symmetry. The reason for this property is also
thought to be the closeness to the antiferromagnetic phase as the lines of
nodes of the superconducting gap allows for the existence of gapless spin excitations
and superconductivity can coexist with spin-density wave fluctuations.  
Very recently, the proximity between these two phases supplemented with the effect of
disorder
 has been the basic ingredient  in a phenomenological model to suggest that {\em colossal}
effects can be expected in the phase diagram of the cuprates \cite{col}.

The fact that the ground-state of the undoped materials is antiferromagnetic
 immediately suggests that the Hubbard model could be a good candidate 
for a microscopic description of these compounds. Instead, the explanation
of the superconducting mechanism in the framework of this model remains 
very controversial. A superconducting solution is found when the model is 
tackled with some many-body techniques\cite{flex1,flex2,flex3,guinea,drazen1,zs96,metzner,honer,hlubina}. 
However, all the numerical works devoted
to search indications of superconductivity in the repulsive Hubbard model have been
negative so far \cite{moreo,assad,trem,zhan}. Another candidate model to provide the basis for the
theoretical investigation of the high T$_c$ superconductors is the $t-J$ model, which
coincides with the Hubbard Hamiltonian in the limit of 
$U \rightarrow \infty$ (corresponding to $J \rightarrow 0$). In the case of 
the $t-J$ model, many-body methods and numerical results seem to be in agreement
concerning the possibility of a superconducting state with 
$d_{x^2-y^2}$ symmetry \cite{kotliar,riera,elbio,mc}. The range of $J$ at which it may
occur is, however, not precisely determined and it is likely
to lie beyond the region where the mapping from the Hubbard model 
is valid. 

This  motivates the study of the $t-J-U$ model
\begin{equation}
H=-t\sum_{\langle ij \rangle, \sigma} 
(c^{\dagger}_{i\sigma} c_{j \sigma} + hc ) + 
J \sum_{\langle ij \rangle } {\bf S}_i {\bf \cdot} {\bf S}_j +
 U\sum_i n_{i \uparrow} n_{i \downarrow} ,
\label{ham}
\end{equation}
which, in addition to the Coulomb repulsion $U$ of the usual Hubbard 
Hamiltonian, contains a nearest-neighbor
 exchange interaction $J$ as the $t-J$ model. In such a way,
we can expect to retain appealing features of both models, like the charge fluctuations
introduced by $U$ but forbidden in the constrained $t-J$ model and the robust
superconducting correlations introduced by the exchange interaction and explore
the interplay between both effects. Beside these heuristic arguments, 
 the $t-J-U$ model is closely related to the extended Hubbard model
with correlated hopping, which has been derived from the three-band extended
Hubbard model as an effective one-band Hamiltonian to describe the low energy
properties of the Cu-O planes of the high-$T_c$ materials \cite{sim}. 
The latter model is characterized by three
different nearest-neighbor hopping amplitudes 
$t_{AA}, t_{AB}, t_{BB}$ depending 
on the occupation of the two sites involved, as well as the usual Coulomb
repulsion $U$,
\begin{eqnarray}
H &=& U\sum_i n_{i \uparrow} n_{i \downarrow} 
-\sum_{\langle ij \rangle, \sigma} (c^{\dagger}_{i\sigma} c_{j \sigma} + hc )
\nonumber \\
& & \{t_{AA} (1-n_{i \overline{\sigma}}) (1-n_{j \overline{\sigma}}) \nonumber \\
& &t_{AB} [(1-n_{i \overline{\sigma}}) n_{j \overline{\sigma}}) +
n_{i \overline{\sigma}}(1- n_{j \overline{\sigma}}) ] \nonumber \\
& & t_{BB} n_{i \overline{\sigma}} n_{j \overline{\sigma}} \}.
\label{corhop}
\end{eqnarray}
Each of the parameters of the above model depends on the parameters of the
original three-band Hamiltonian and there is a degree of uncertainty 
in their precise values. The relevant property is that reasonable estimates
indicate that $t_{AB}$ is larger than the other two \cite{sim}. 
The analysis of the
different hopping processes in (\ref{corhop}) reveals that the one
driven by $t_{AB}$ mediates antiferromagnetic 
correlations. In particular, in the strong coupling limit 
$U>>t_{AA},t_{AB},t_{BB}$,
the exchange interaction obtained by treating  (\ref{corhop}) with second
order perturbation theory is $J=4t_{AB}^2/U$. 
Furthermore, when this process is suppressed, antiferromagnetic
correlations are completely eliminated and a metal-insulator transition can take
place at finite $U$ in the half-filled system \cite{esol1,esol2,tuv}.
For weak coupling, the Hamiltonian (\ref{corhop}) can
be treated with mean-field Hartree-Fock and BCS-like techniques in 2D \cite{lidw} and 
with operator product expansion and bosonization in 1D \cite{trip}. It turns 
out that, close to half-filling, the predicted phase diagram is equivalent to 
that obtained starting from an effective $t-J-U$ model \cite{note}. 
 In summary, the relevance of the $t-J-U$ model
to describe the Physics of the cuprates can be also supported by its 
closeness to the correlated Hubbard model derived from the more detailed
three-band Hubbard model for the Cu-O planes.  

Coming back to the phenomenology of the superconducting cuprates, the $t-J-U$ model
provides the framework for recent suggestions based on a combined order parameter
with superconducting and bond-density-wave components with $d_{x^2-y^2}$ symmetries
to explain the intriguing pseudogap phenomena \cite{laugh}. In general,  
extended Hubbard models in low dimensions are of interest in the context of
high-$T_c$ materials \cite{hir,paco,mar} and also to understand
the rich structure observed in the phase diagrams of organic materials like
(TMTSF)$_2$PF$_6$, and (TMTSF)$_2$ClO$_4$ 
\cite{tuv,trip,jap,douc,dolci,tsu,orga1,orga2,orga3,orga4}, which can not
be explained on the basis the local Coulomb interaction of the usual Hubbard model.   

The $t-J-U$ model has an explicit effective attraction mediated by $J$ that
enables a superconducting solution even within a simple BCS-like 
description. Numerical studies of this model in ladders indicate that
superconducting correlations with $d_{x^2-y^2}$ symmetry are enhanced
in comparison to those of the usual Hubbard model \cite{daul}. 
More recently, an analysis based on the Landau Fermi liquid picture has been
adopted to argue that the pairing interaction mediated by $J$ combined by
a strong renormalization of the effective density of states caused by $U$
results in a significant enhancement of the superconducting order parameter
while quantum Monte Carlo simulations based on a variational wave function
with a BCS structure support this picture \cite{ple}. Investigations
of the weak coupling phase diagram at van Hove fillings
 of the two dimensional version of this model 
supplemented by a nearest-neighbor Coulomb repulsion and next-nearest neighbor
hopping amplitude have also been reported \cite{kat}.

The aim of this work is to investigate
the two-dimensional phase diagram of the $t-J-U$ 
model with two complementary techniques: exact diagonalization (ED) of a small
cluster with $4\times4$ sites and the one-loop renormalization group (RG) 
technique presented in Refs. \cite{drazen1,drazen2}.
The latter are expected to provide reliable indications of the main instabilities of the
Fermi liquid in the limit of weak interactions. 
ED is an unbiased method that consists in the exact calculation 
of the ground state wave function, allowing for the direct evaluation of the relevant correlation
functions  but has the drawback that the size of the clusters that are amenable
to be treated is small. In spite of that limitation,
 in the case of the $t-J$ model, the conclusions 
 based on numerical results on  such small clusters are 
among the most robust ones regarding superconductivity \cite{elbio}.
It is therefore interesting to analyze the predictions of these two methods
and to compare them with previous results.

The article is organized as follows, in section II we provide some technical 
details on the methods we employ. In particular, we describe the procedure
 followed
to include self energy corrections in the RG treatment.
An important reference is the behavior of the relevant susceptibilities 
evaluated with this RG procedure 
for the usual Hubbard model. 
This is  discussed in Section III. 
In sections IV and V we present the results for the $t-J-U$ model using RG and ED, respectively. 
Section VI is devoted to summary and conclusions.

\section{Technical details}  
\subsection{Renormalization group method}
The basic hypothesis of the renormalization group method 
is that the normal state is well described by the effective action for 
quasiparticles near the Fermi surface. 
The description of the many body problem is done in terms of the two-body effective interaction
$U_l(\theta_1,\theta_2,\theta_3)$ and the  quasiparticle weight $Z_l(\theta)$.
In the Wilson's renormalization scheme,  $U_l(\theta_1,\theta_2,\theta_3)$
is the effective interaction between electrons {\em within} the ring 
$\pm \Lambda$ around the Fermi surface. This interaction is renormalized 
by  the scattering processes involving all electrons {\em outside} 
the ring $\pm \Lambda$. A physical interpretation
of the cutoff $\Lambda$ is that it plays the role of an effective temperature or the
experimental probe frequency.
In  previous versions of RG \cite{drazen1,honer,metzner,kat}, it was assumed 
$Z_l(\theta)$=1. In what follows, we summarize the improved RG method 
of Ref. \cite{drazen2} which also considers self-energy corrections by taking
into account the renormalization of the quasiparticle weight. 

\begin{figure} 
\centerline{\psfig{file=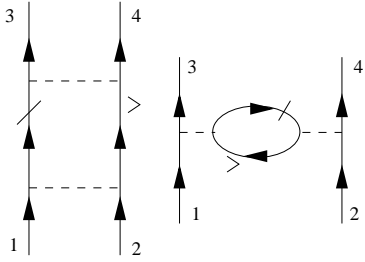,width=0.29\textwidth,angle=0}} 
\hfill 
\caption{\small 
   From the left: Particle-particle (Cooper) and particle-hole 
(density-wave)
    differential bubbles $\beta _{pp}(\theta_1,\theta_2,\theta_3)$ and
$\beta _{ph}(\theta_1,\theta_2,\theta_3)$. Propagators corresponding to
degrees of freedom at the Fermi surface and outside the ring $\pm \Lambda$
are indicated with a bar and a $>$, respectively.} 
\label{bete} 
\end{figure}

An important issue to note is 
the fact that $U_l(\theta_1,\theta_2,\theta_3)$ and $Z_l(\theta)$ depend only on the
angles $\theta_i$ that parametrize the positions of the particles on the Fermi surface.
This is justified by a simple power counting which tells us that
only the angular dependence of the effective interaction
is marginal (or marginally-relevant) and only terms up to linear in energy are to be kept in the 
renormalization of the angle-dependent self-energy \cite{drazen2}.
In the renormalization group procedure, these functions
are continuously renormalized as the energy 
cutoff, parametrized by the scale $l$ as $\Lambda=8t \exp{(-l)}$ is reduced.
The ensuing equation for the evolution of $U_l$ within the one-loop 
renormalization group scheme has
the following structure:
\begin{equation}
\frac{\partial{U_l}}{\partial{l}}
=\beta _{pp}\{ U_l,U_l\} +2{\beta}_{ph}\{ U_l,U_l\}-
{\beta}_{ph}\{ U_l,XU_l\}-{\beta}_{ph}\{ XU_l,U_l\}-X{\beta}_{ph}\{ XU_l,XU_l\},
\label{RG_U}
\end{equation}
where $X$ is the exchange operator defined as $XU(1,2,3)=U(2,1,3)$.
One must remember that (\ref{RG_U}) is a {\em functional} flow equation, i.e. $U_l$ and
all terms on the right-hand side depend on three angles
$(\theta_1,\theta_2,\theta_3)$. Particle-particle (Cooper) and 
particle-hole (density-wave) differential bubbles 
$\beta _{pp}$ and $\beta _{ph}$ are shown on fig.\ref{bete}.  
To solve numerically Eq.(\ref{RG_U}) we discretize the 
angle $\theta$ which defines the so-called $N$-patch model. 
The function $U_l(\theta_1,\theta_2,\theta_3)$ is then
represented by a set of coupling constants labeled by three discrete indexes.
For the $t-J-U$ model, the initial condition is
\begin{equation}
{U}_{l=0}(\theta_1,\theta_2,\theta_3)=
\left\{
U-\frac{J}{4}[\cos (x_3-x_1)+\cos (y_3-y_1)]- 
\frac{J}{2}[\cos (x_3-x_2)+\cos (y_3-y_2)]
\right\} \; ,
\label{ic}
\end{equation}
where $(x_i,y_i)$ are components of the wave vector ${\bf k}_i$, corresponding 
to the angle $\theta_i$ at the noninteracting Fermi surface.
All coupling constants are found to diverge at the same critical scale $l_c$
like
\begin{equation}
{U}_l(\theta_1,\theta_2,\theta_3)\rightarrow 
\frac{\tilde{U}(\theta_1,\theta_2,\theta_3)}{l_c-l}\; ,
\label{div}
\end{equation}
 where the weights $\tilde{U}$
are model dependent constants. 
This type of solution
is called the fixed-pole solution in contrast to the mobile-pole solution,
where different coupling constants  diverge at different critical scales.
For realistic systems, where the initial coupling is not extremely small, only 
fixed poles are relevant \cite{doucot_00}.
The critical scale $l_c$ depends on the bare coupling constants $U$ and $J$ and on the
band filling parametrized by the chemical potential. The critical cutoff
$\Lambda _c=8t\exp{(-l_c)}$ appears to be the
characteristic temperature of the  model. The most precise 
non-restrictive interpretation of $\Lambda _c$ is that at this energy
electrons start to build bound states. Namely, the  poles in Eq.\ref{div} 
are the two-particle propagator poles, indicating the on-set of 
bound states. These bound states can be of charge, spin or superconducting
kind and all of them  renormalize the one-particle weight. 
This renormalization is angle dependent and its evolution is described by the following expression
\begin{equation} \label{RG_polelog}
\partial _l\log{Z}_l(\theta)=\frac{1}{(2\pi)^2}\int d\theta'\; {\cal J}
(\theta',-\Lambda)\eta_l(\theta,\theta')\; ,
\end{equation}
with the initial condition ${Z}_l(\theta)=1$. The function
${\cal J}
(\theta,\epsilon )$ is the angle dependent density of states at the energy 
$\epsilon$ (measured from the Fermi level).
The quantity $\eta_l(\theta,\theta')$ contains particle-particle (pp) and 
particle-hole (ph) contributions
\begin{equation} \label{eta}
\eta_l(\theta,\theta')\equiv (2X-1)\beta _{pp}\{ U_l,U_l\} 
+2\beta _{ph}\{ XU_l,XU_l\} +2\beta _{ph}\{ U_l,U_l\} -\beta _{ph}\{ U_l,XU_l\}-
\beta_{ph}\{ XU_l,U_l\} 
\end{equation}
with all terms on the right-hand side taken with external legs
$(\theta_1,\theta_2,\theta_3) =(\theta,\theta',\theta')$.
The interaction inserted in all beta functions obeys
the scaling equation (\ref{RG_U}). 
The relation between interaction
$U_l$ and the usual physical interaction $\Gamma$ contains the rescaling 
$Z$-factors, \cite{drazen2}:
\begin{equation} \label{resc_U}
\Gamma_l(1,2,3)=[Z_l(1)Z_l(2)Z_l(3)Z_l(4)]^{-1/2}U_l(1,2,3)\; .
\end{equation}
 To find out which correlations are relevant and candidates for 
order parameter
 we must allow the theory to choose 
between all possible 2-particle correlations.
For this reason we have to follow the renormalization of several
angle-resolved correlation functions.
The superconducting correlation function $\chi
^{SC}_l(\theta_1,\theta_2)$ measures  correlations between 
the Cooper pairs 
$(\theta_1,\theta_1+\pi)$ and $(\theta_2,\theta_2+\pi)$, all states being at
the Fermi surface. 
The antiferromagnetic correlation function 
$\chi ^{AF}_l(\theta_1,\theta_2)$ correlates two nested electron-hole pairs
$c^{\dagger}_{{\bf k}(\theta_1)}{\bf \sigma}c_{{\bf
k}(\theta_1)+(\pi,\pi)}$ and 
$c^{\dagger}_{{\bf k}(\theta_2)}{\bf \sigma}c_{{\bf k}(\theta_2)
+(\pi,\pi)}$.
The charge density wave correlation function $\chi ^{CDW}_l(\theta_1,\theta_2)$
correlates the nested charge-like electron-hole pairs.
To get the renormalization group flow of all correlation functions we
follow the procedure given in Ref.\cite{drazen1} but dressing the
electronic propagators with
$Z$--factors as in Ref.\cite{drazen2}.  We get 
\begin{equation} \label{flow_chi}
\dot{\chi}_l^{\delta}(\theta_1,\theta_2)=\frac{1}{Z_l(\theta_1)Z_l(\theta_2)}
\oint d\theta\; 
\tilde{z}_{l}^{\delta}(\theta_1,\theta ) D_l^{\delta}(\theta ) 
\tilde{z}_{l}^{\delta}(\theta
,\theta_2) \; .
\end{equation}
The function $D_l^{\delta}(\theta)$ writes
\begin{equation} \label{D-SC}
D_l^{SC}(\theta)=\frac{1}{2}\sum _{\nu=+,-}{\cal J}(\nu\Lambda(l),\theta)
\end{equation}
for the SC channel and
\begin{equation} \label{D-DW}
D_l^{AF}(\theta)=\frac{1}{2}\frac{{\cal J}(-\Lambda(l),\theta)
}{1+|\mu|/\Lambda(l)}\; ,
\end{equation}
for the AF channel where only the negative shell $(\nu=-1)$
contributes to the flow. 

The flow of the quantity
$\tilde{z}_{l}^{\delta}(\theta_1,\theta )$ that has the role of a triangular vertex
writes:
\begin{equation} \label{flow_z}
\left[ \partial _l-\eta(\theta_1)-\eta(\theta_2)\right] 
\tilde{z}_l^{\delta}(\theta_1,\theta_2)=-\oint d\theta \; 
\tilde{z}_{l}^{\delta}(\theta_1,\theta ) D_l^{\delta}(\theta )
V_{l}^{\delta}(\theta ,\theta_2) \; .
\end{equation}
The meaning of $\tilde{z}_l^{\delta}(\theta_1,\theta_2)$ is that
$$
\tilde{z}_l^{\delta}(\theta_1,\theta_2)\equiv Z_l(\theta_1)
{z}_l^{\delta}(\theta_1,\theta_2)Z_l(\theta_2)
$$
so that the initial conditions for $\tilde{z}_l^{\delta}$ and for
${z}_l^{\delta}$ are the same:
\begin{equation} \label{z-incond}
z_{l=0}^{\delta}(\theta_1,\theta_2)=\delta _D(\theta_1-\theta_2),
\end{equation}
where $\delta _D$ is the Dirac function.
Initial conditions for susceptibilities are:
\begin{equation} \label{chi-incond}
\chi_{l=0}^{\delta}(\theta_1,\theta_2)=0.
\end{equation}
After  discretization we integrate numerically
equations (\ref{flow_chi}) and (\ref{flow_z}).
The relevant susceptibility in each channel is the dominant eigenvalue
of the angle-resolved correlation function. 
The corresponding eigenvector
determines the angular dependence of the order parameter.

\subsection{Exact diagonalization}
This method consists in the exact computation,
by recourse to Lanczos algorithm,
 of the ground state (GS) wave function
 of the model Hamiltonian (\ref{ham}) on a small cluster. We consider a cluster
 containing $4\times 4$ lattice sites.

\begin{figure}
 \epsfxsize=3.5in
  \epsffile{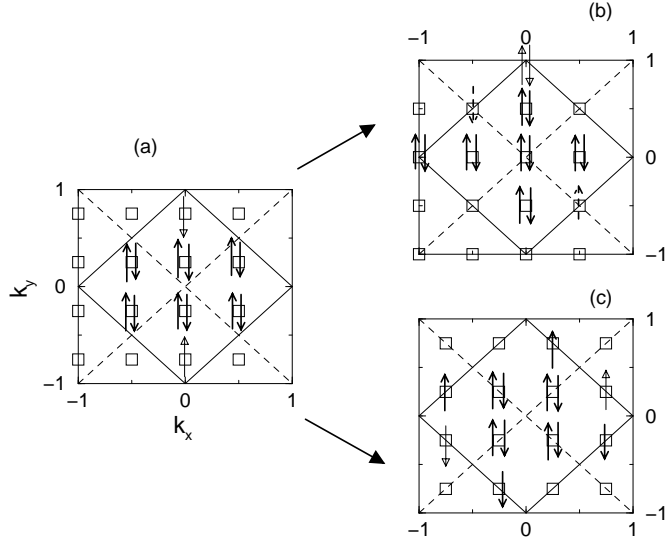}
  \caption{Reciprocal lattice of the $4\times 4$ cluster assuming (a) mixed boundary
conditions, (b) periodic boundary conditions and (c) antiperiodic boundary conditions.
Ground state configurations corresponding to the fillings $n=0.75$ (thick arrows)
and $n=0.875$ (adding the thin arrows to the previous case) are also 
sketched.}
\label{fig2}
\end{figure}
 
Due to the finite size of the cluster, its reciprocal lattice  
contains just some few ($N=16$) points. As several instabilities are expected to be 
competitive in this model,
the rough mesh of available ${\bf k}$-vectors can produce an important
bias and lead to an artificial enhancement of some kind of correlation.
The only resort to minimize this drawback is to consider different kinds
of boundary conditions, which is equivalent to consider different 
choices of the 16 ${\bf k}$-points. An arbitrary choice of boundary conditions
breaks the point group symmetry of the original lattice. 
Exceptions are
the periodic (P) and antiperiodic (AP) boundary conditions, which lead to the
mesh indicated in Figs.\ref{fig2}b and c, respectively. Mixed (M) boundary
conditions
(periodic in one direction and antiperiodic in the other) lead to the
pattern depicted in Fig.\ref{fig2}a. The latter breaks some of the symmetries of the 
$C_{4v}$ group of the original lattice, the corresponding point
group being $C_{2v}$.  Typically, these three choices of boundary conditions
are the ones leading to the lowest energy. In our study, we compute
the GS of (\ref{ham}), considering all the three possibilities above 
mentioned within the subspaces corresponding to the different one-dimensional
representations of the point group and total ${\bf k}=(0,0)$.
 In the noninteracting system, it is easy to see that closed-shell 
configurations are those leading to the lowest energy. For some densities of
 particles
this condition is, however, not possible to be fulfilled for any choice
of the boundary conditions  and the
GS is degenerate. The interactions normally lift most of
the degeneracies. In some cases, it is observed that when the
interactions overcome some particular value, a change is produced in the BC
leading to the lowest energy. The latter effect
is an indication 
that interactions lead to some qualitative change in the behavior 
of the GS. This is, of course, a mere finite size effect 
but provides a valuable information, since it reflects
that the system prefers a
change in the population of available ${\bf k}$-points
in order to take advantage of the interactions and thus lower the energy.

To investigate the superconducting correlations in the GS
it is useful to study the behavior of the pair correlation function (PCF)
\begin{equation}
P({\bf r})\;=\;\frac{1}{N} \sum_i
\langle \Psi_0|
\Delta _{\alpha }^{\dagger }({\bf R}_i + {\bf r})
\Delta _{\alpha}({\bf R}_i) |\Psi_0 \rangle ,  \label{pcf}
\end{equation}
where $|\Psi_0 \rangle$ is the ground-state wave function while
 $\Delta _{os}^{\dagger }({\bf R}_i)=c_{i\uparrow }^{\dagger }c_{i\downarrow}^{\dagger }$ 
for on-site $s$ pairing, $\Delta _{\alpha }^{\dagger}({\bf R}_i)=\sum_{\delta }
f_{\alpha }(\delta )[c_{i+\delta \uparrow }^{\dagger}c_{i\downarrow }^{\dagger }-
c_{i+\delta \downarrow }^{\dagger }c_{i\uparrow}^{\dagger }]/\sqrt{8}$, 
with $f_{es}(\delta )=1$ for extended $s$ pairing,
and $f_{d}(\delta )=1$ ($f_{d}(\delta )=-1$ ) when $\delta =\pm (1,0)$ 
($\delta =\pm (0,1)$) for $d_{x^{2}-y^{2}}$ pairing. 
This function is normalized in such a way
that $|\Delta _{\alpha }^{\dagger }(i)|\Psi_0\rangle |^{2}=1$.
A superconducting state with pairs of a given symmetry is expected to have
sizable correlations between pairs far separated by arbitrary large distances. In the case of
the $4 \times 4$ cluster, the largest available distance is ${\bf r}=(2,2)$.
The  PCF between pairs separated by this maximum distance is denoted $P_m$. As even in the
noninteracting limit  the PCF can be finite, we interpret 
an enhancement of the corresponding PCF {\em relative to its value  at $U=0, J=0$}
as an indication of the superconducting instability.  

To study the spin-density-wave (SDW) correlations it is useful to compute the
spin-spin correlation function,
\begin{equation}
S({\bf r})\;=\;\frac{1}{N} \sum_i  
\langle \Psi_0| S^z( {\bf R}_i + {\bf r}) S^z( {\bf R}_i )|\Psi_0 \rangle ,  
\label{scf}
\end{equation}
and to analyze the Fourier transform
\begin{equation}
S({\bf k})\;=\;\frac{1}{N} 
\sum_i e^{i {\bf k \cdot R}_i}  S({\bf r}),
\end{equation}
which provides information on the nature of the correlations between spins. 

\section{Renormalization-group results for  Hubbard model with repulsive and 
attractive interaction}
In this section we present results for the usual Hubbard model obtained by the RG method
described in the previous section. The motivation is twofold. 
For the case of the $U>0$ model, it was shown  in Ref. \cite{drazen2} 
that self-energy corrections included in the renornalization of $Z(\theta)$
are important to predict the antiferromagnetic instability at half-filling in 2D and
to recover the correct expression for the jump at the Fermi points in 1D but the
behavior of the susceptibilities away from half-filling has not been analyzed so far.
On another hand, for $U<0$ the model is a paradigmatic example of a superconductor 
and it is, therefore, an important reference point to analyze the behavior of the 
superconducting correlations. The results shown correspond to a discretization of 32 patches.

\begin{figure}
 \epsfxsize=3.5in
\includegraphics[width=65mm,angle=-90]{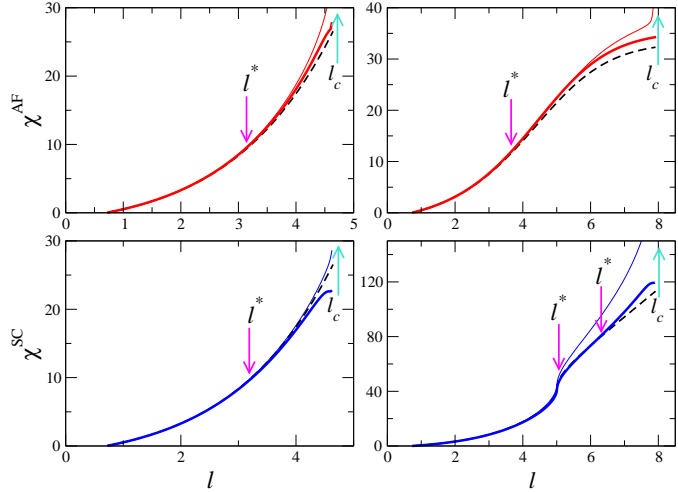}
  \caption{(Color online) AF (upper panels) and $d_{x^2-y^2}$  SC (lower panels)
 susceptibilities as functions of the scale $l$ for Hubbard model with $U=1.6 t$ and 
densities determined by 
$l_{\mu}=\infty$ (left-hand panels) and $l_{\mu}=3$ (right-hand panels), corresponding,
respectively, to half-filling (parquet regime) and high doping (within the BCS regime).
Thick and thin lines correspond to results with and without self-energy corrections, respectively.
The noninteracting particle-particle and particle-hole susceptibilities are
shown in dashed lines. The critical scale $l_c$ as well
as the scale $l^*$ at which the susceptibility departs from the behavior of the
noninteracting one are indicated with magenta and cyan arrows, respectively.} 
\label{posu}
\end{figure}
When applied to the usual Hubbard model with repulsive
interactions ($U>0$), the RG without
considering self-energy corrections \cite{drazen1,metzner,honer}, predicts 
that at some critical scale $l_c$, the superconducting
susceptibility with $d_{x^2-y^2}$ symmetry together with AF susceptibility
diverge within the parquet regime ($|\mu|<\Lambda$), even at half-filling. 
This feature is expected to be
an artifact of the approximation since for $n=1$ umklapp processes are
active and they are expected to drive the system toward an insulating AF state.
In Ref. \cite{drazen2} it was shown that, due to the renormalization of the
quasiparticle weight,  
the SC susceptibility is suppressed below its noninteracting value, 
while the
AF susceptibility remains enhanced, although weakly,  
relative to the noninteracting particle-hole one.

 Fig. \ref{posu} shows the behavior for the different susceptibilities as functions 
of the scale $l$, at different densities defined by the 
chemical potential $\mu=8t \exp(-l_{\mu})$. The susceptibilities with and without 
self-energy corrections are respectively plotted in thin and thick lines.
The two left-hand panels correspond to half-filling ($l_{\mu}=\infty$)
and summarize the results of Ref. \cite{drazen2}. Without self-energy corrections
both SC and AF susceptibilities diverge at $l_c$, AF being dominant.
Self-energy effects suppress the divergence, decreasing the SC susceptibility bellow its
noninteracting value. The AF susceptibility is also renormalized by self-energy
effects but remains larger than the noninteracting one.
The scale $l^*$ (see Fig. \ref{posu}) at which
the AF susceptibility begins to depart from the behavior of the noninteracting 
one is, however, not affected by the self-energy corrections.    
This kind of behavior extends along 
the parquet regime, defined by the condition $|\mu|<\Lambda$.

The region at lower densities such that $|\mu|>\Lambda$ is usually
called BCS regime because only Cooper channel has logarithmic contributions to the
effective interaction. The two right-hand panels of Fig. \ref{posu} show how
in this regime
both AF and SC susceptibilities remain slightly larger than their
noninteracting values when self-energy corrections are taken into account.

These results need some discussion. Up to now it was rather widely accepted that
the 2D Hubbard model has a $d_{x^2-y^2}$ SC state away from half-filling.
On the contrary results on the right-hand panels of Fig. \ref{posu} do not 
convincingly  indicate 
a strong superconducting instability. The SC correlation functions being only weakly enhanced 
and the scale $l^*$ being strongly renormalized towards high values when 
self-energy effects are taken into account (see right-hand lower panel of
Fig. \ref{posu}). This indicates that the energy $\Lambda^*=8t \exp{(-l^*)}$
at which superconducting correlations begin to manifest themselves decreases when self-energy effects
are considered.    
At $\Lambda_c$ there is a creation of bound states because the
4-points vertex (\ref{div}) has poles. These bound states are the gapless modes that destroy the one-particle coherence via self-energy corrections and the result is that the phase transition is 
suppressed, probably to some finite lower energy.

\begin{figure}
 \epsfxsize=3.5in
\includegraphics[width=80mm,angle=-90]{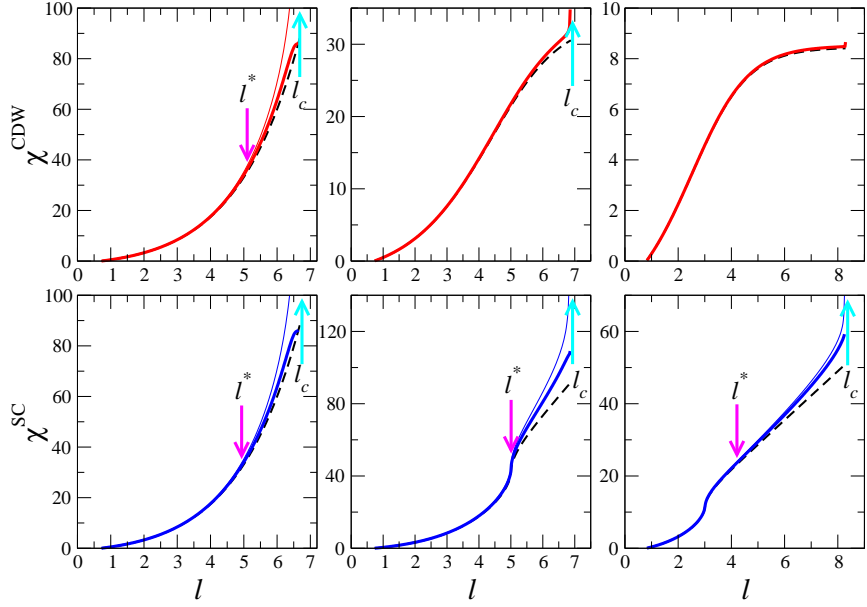}
  \caption{(Color online) CDW (upper panels) and $s$-wave SC (lower panels) 
susceptibilities as functions of the 
scale $l$ for $U=-0.8t$ 
and densities determined by 
$l_{\mu}=\infty, 5, 3$, corresponding to half-filling and two densities within the
BCS regime (from left to right). 
Other details are the same as in Fig. \ref{posu}.} 
\label{negu}
\end{figure}
To justify the above interpretation we calculated for comparison the RG flow of the
attractive Hubbard model, for which the instabilities are better known.
The $U<0$ model    
is a paradigmatic example of $s$-wave type superconductor and its
phase diagram has been investigated in detail by mean-field and numerical
techniques. Unlike the repulsive counterpart, predictions by different
methods agree about the main instabilities expected in
its phase diagram \cite{scal,ran}:   
At weak coupling, the explicit local negative 
interaction leads to BCS-like superconductivity away from half-filling. For
$n=1$, local CDW is believed to be degenerate with $s$-wave superconductivity in the 
GS \cite{scal}.
In 1D, such degeneracy is exact due to symmetry reasons. This is because
the repulsive model has dominant AF correlations
with a power law decay. Since no breaking of the SU(2) symmetry is possible, the behavior of 
the correlations
in any of the spacial direction must behave in the same way. 
On another hand, there is an exact 
transformation
$c^{\dagger}_{i \uparrow} \rightarrow (-1)^i c_{i \uparrow}$, 
$c^{\dagger}_{i \downarrow} \rightarrow c^{\dagger}_{i \downarrow}$, which maps the 
repulsive model
into the attractive one, while maps the degenerate $z$ and $x,y$ components of the dominant SDW
correlations of the $U>0$ model to the CDW and superconducting ones, respectively, 
of the $U<0$ counterpart. The above reasoning can also be extended to the 2D case
provided that no symmetry breaking in the ground state of the $U>0$ model takes place.  
Results obtained with RG for the attractive model in 2D shown in Fig \ref{negu} are 
in very good agreement with the description provided by numerical 
methods \cite{scal,ran}.
At half-filling ($l_{\mu}=\infty$, see left-hand panels of Fig. \ref{negu}), 
the most remarkable feature
is the degeneracy observed between CDW and SC susceptibility with local
$s$-wave symmetry, which remain slightly  larger than the noninteracting one when
self-energy corrections are considered. This should be rather expected 
since the present method provides a description of the normal state and only the on-set 
of the instability towards the symmetry broken state is captured.
Bellow half-filling ($l_{\mu}=5, 3$, cf. middle and right-hand panels of Fig \ref{negu}), 
CDW susceptibility becomes weaker, approximately equal to the noninteracting
one, while the superconducting susceptibility
becomes more enhanced. Self-energy effects suppress the divergence of the different instabilities
but do not renormalize the scale $l^*$.

Bellow half-filling, the $s$-wave type SC susceptibility
 is clearly larger than the noninteracting particle-particle one. As expected, the CDW remains non renormalized.
Compared to the repulsive Hubbard model (Fig.\ref{posu}), the enhancement of SC correlations is more convincing
 and in the present case we can indeed assign
the divergence at $l_c$ to the on-set of the $s$-wave superconductivity.
Altogether, the attractive Hubbard model away from half-filling shows strong tendencies 
towards the $s$-wave superconductivity even when the self-energy corrections are taken into account.
This is in contrast to the repulsive case where the self-energy effects
 have a more pronounced effect against
the $d_{x^2-y^2}$ superconductivity. 
The reason for this behavior is rather simple: the $s$-wave superconductivity
 in the attractive Hubbard model 
is an effect of first order in $U$, while the self-energy corrections are of second order.
As long as the coupling is weak the BCS like instability is a good approximation.
Formally this means that the mean-field and the Kosterlitz-Thouless transitions are close to
one another. In the case of the repulsive interaction the $d_{x^2-y^2}$ superconductivity and the 
self-energy corrections are 
both of  second order in $U$, so that by decreasing $U$ one cannot promote only
superconductivity and make the fluctuations subdominant.

\section{Renormalization-group results for the $t-J-U$ model.}
\begin{figure}
\includegraphics[width=65mm,angle=-90]{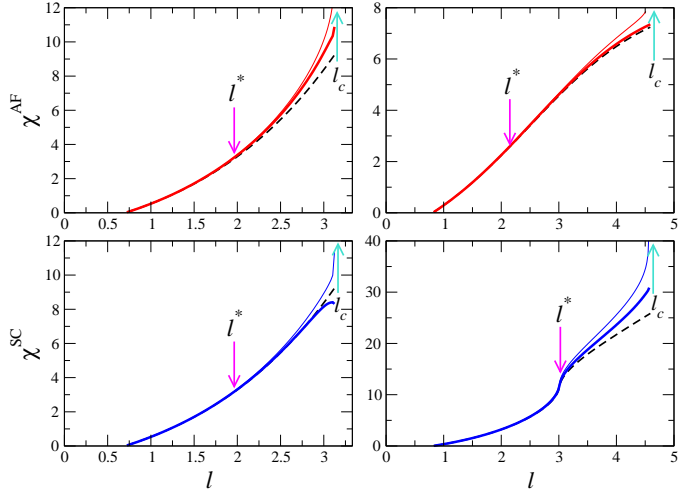}
  \caption{
(Color online) AF (upper panels) and $d_{x^2-y^2}$  SC (lower panels)
 susceptibilities as functions of the scale $l$ for Hubbard model with $U = J = 1.6 t$ and 
densities determined by 
$l_{\mu}=\infty$ (left-hand panels) and $l_{\mu}=3$ (right-hand panels), corresponding,
respectively, to half-filling (parquet regime) and high doping (within the BCS regime).
 Other details are the same as in Fig. \ref{posu}. }
\label{figuj}
\end{figure}
We present results for the relevant susceptibilities in the $t-J-U$ model in Fig. \ref{figuj}.
The left-hand panels correspond to half-filling and the behavior is similar to that of the
Hubbard case. Namely, SC susceptibility with $d_{x^2-y^2}$ symmetry
evolves to values below of the noninteracting one,
while AF susceptibility remains higher than its noninteracting value, indicating that at $n=1$, the
system flows towards an AF insulating state.
Upon doping the $d_{x^2-y^2}$ superconductivity gets progressively stronger and becomes the dominant 
instability at about the crossover line $\Lambda = \mu$, just as in the Hubbard model.
The right-hand panels show a typical flow of the susceptibility in the BCS regime
($\Lambda < \mu$) where the dominant correlations are of superconductivity type.

The above picture shares some features with the behavior observed in the Hubbard model, discussed
in the previous section. Important issues to highlight are:
 (i) At half-filling the behavior of the relevant susceptibilities is very similar to that of
the repulsive Hubbard model. However, 
the scale $l_c$ at which the on-set of the AF instability takes place as well as the scale $l^*$
are smaller than for the $J=0$ case. The susceptibility is also significantly larger than the 
noninteracting one, even when self-energy corrections are included in the RG procedure. 
These features indicate  that $J$ contributes to increase the
AF correlations and the Neel temperature. (ii) At higher dopings, within the BCS regime,
AF susceptibility coincides with the noninteracting one while SC correlations become 
significantly enhanced. This is in contrast to the behavior of the repulsive model 
(cf. Fig. \ref{posu}) where SC correlations are only weakly enhanced. Instead, the
behavior on the right panels of Fig. \ref{figuj} resembles the one of the attractive Hubbard model,
if we associate $d_{x^2-y^2}$ SC and AF susceptibilities in Fig.  \ref{figuj} respectively to $s$-wave SC and 
CDW susceptibilities on middle and right panels of Fig. \ref{negu}. 
Also note that, as in the attractive model, the scale
$l^*$ remains unaffected by self-energy effects. In addition,
the scale $l_c$ at which the on-set of the superconducting instability 
is observed is small in the $t-J-U$ model, implying a high critical temperature.
We have carried out a similar analysis for other values of the parameters $J$ and $U$ and found
that the symmetry of the dominant superconducting correlations is always $d_{x^2-y^2}$.     
 We have verified the reliability of these results upon increasing number of patches up to 64 patches.
The reason for the robustness of the SC correlations 
is in the fact that the $J$  interaction has an attractive $d_{x^2-y^2}$ SC component,
so that  superconductivity exists already at first order of
$U_l$ while the fluctuations are only subdominant, 
just as in the attractive Hubbard case.

\begin{figure}
\includegraphics[width=65mm,angle=-90]{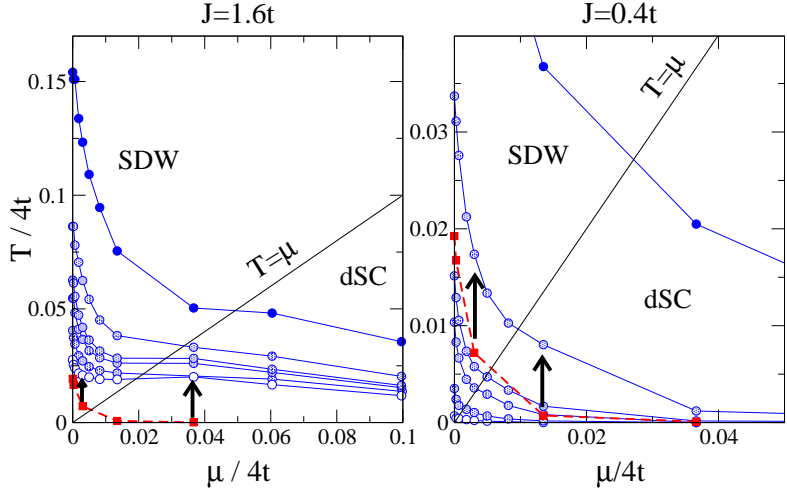}
\caption{ (Color online) Phase diagram showing combined action of $U$ and $J$ interactions.
The ``critical temperature'' is defined as $T_c \sim \Lambda_{c}=8t \exp(-l_{c})$ (see text).
Left and right panels correspond to $J=1.6t$ and $J=0.4t$, respectively. 
Plots in  blue circles correspond to $U=0, 0.4t, 0.8t, t, 1.6t, 3.2t$.
The ones corresponding to the lowest and highest $U$ are drawn in open and dark filled symbols, respectively.
The plot in red squares corresponds to the usual Hubbard model ($J=0$ and $U=1.6t$). 
 The line $T=\mu$ separates the regions where the SDW and $d_{x^2-y^2}$ superconducting (dSC) instabilities are the dominant ones.}
\label{pduj}
\end{figure}

In order to have a more quantitative representation of the role played by both interactions we 
present the phase diagram in Fig. \ref{pduj}. The ``critical temperature''
plotted in the figure is defined as $T_c \sim \Lambda_{c}$ where $\Lambda_{c}=8t \exp(-l_{c})$. 
Results for the repulsive Hubbard model are also shown for comparison.
By comparing the plots for $J=0, U=1.6t$ and $J=1.6t, U=0$ in the left panel and
for $J=0, U=1.6t$ and $J=0.4t, U=1.6t$ in the right panel (the arrows are drawn to ease the reading), 
it is clear that  $J$ is a remarkably 
efficient  mechanism to drive AF close to half-filling and $d_{x^2-y^2}$ superconductivity
in the doped system. Another important feature
is that at fixed $J$, the effect of $U$ is to increase the ``critical temperature''. This
means that the two interactions are not competitive but, instead, cooperate to increase the
strength of antiferromagnetic and superconducting correlations.

\section{Exact-diagonalization results for the $t-J-U$ model in the $4 \times 4$ cluster.}
The results of the previous section suggest that the combined effect of the
 interactions $J$ and $U$ drives superconductivity with  $d_{x^2-y^2}$ symmetry
in the doped system, 
leading to a significant enhancement of the superconducting 
correlations. 
 As in the case of 
the pure Hubbard model numerical methods fail to detect the tendency towards 
$d_{x^2-y^2}$-superconductivity, it is interesting to analyze
the case of finite $J$. 

We show bellow results obtained by following
the strategy explained in Section II. We computed the
exact GS energy and wave function in a $4 \times 4$ cluster and calculated the correlation functions
between pairs with local and extended $s$-wave and $d_{x^2-y^2}$ symmetries.
We focused our attention on the study of three different
fillings ($n=0.625, 0.75, 0.875$), corresponding to
$N=10,12,14$ particles in the cluster.

Let us begin with the analysis of the boundary conditions leading to the lowest energy.
For the case $n=0.625$,
the GS is obtained for PBC. In the
noninteracting limit, it corresponds to a closed-shell configuration 
 and for all the explored values of the interactions,
it lies in the subspace associated to the
representation of the point group with $s$-wave like character.
The
${\bf k}$-points lying on the Fermi surface of the noninteracting
system are $(\pi/2,0)$ and the symmetry related points. For these
points, the structure factors $f_{es}({\bf k})=\cos(k_x)+\cos(k_y)$
and $f_d({\bf k})=\cos(k_x)-\cos(k_y)$, corresponding
to BCS gaps with extended $s$ and $d_{x^2-y^2}$-wave symmetries have
exactly the same strength $|f_d|=|f_{es}|=1$.

\begin{table}
\begin{tabular}{cccccccccc}
\tableline
\tableline
$U$ & $ J=0$ & $J=0.25$ & $J=0.5$ & $J=0.75$ & $J=1$ & $J=1.5$ & $J=1.75$ & $J=2$ \\
\tableline
0  & -1.5607 & -1.5940 & -1.6298 & -1.6681 & -1.7092 & -1.8006 & -1.8514 & -1.9057\\ 
2  & -1.3233 & -1.3651 & -1.4101 & -1.4584 & -1.5100 & 
-1.6239 & -1.6864 & -1.7526\\   
4  & -1.1607 & -1.2094 & -1.2616 & -1.3176 & -1.3772 & -1.5075 & -1.5803$^{*}$ & -1.6619$^{*}$\\  
6  & -1.0516 & -1.1045 & -1.1615 & -1.2225 & -1.2874 & -1.4291$^{*}$ & -1.5125$^{*}$ & -1.5975$^{*}$\\    
8  & -0.9774 & -1.0326 & -1.0924 & -1.1565 & -1.2247 & -1.3773$^{*}$ & -1.4630$^{*}$ & -1.5503$^{*}$\\  
10 & -0.9255 & -0.9818 & -1.0430 & -1.1089 & -1.1792 & -1.3385$^{*}$ & -1.4258$^{*}$ & -1.5145$^{*}$\\ 
12 & -0.8878 & -0.9445 & -1.0065 & -1.0735 & -1.1451 & -1.3087$^{*}$ & -1.3969$^{*}$ & -1.4867$^{*}$\\ 
14 & -0.8596 & -0.9162 & -0.9786 & -1.0463 & -1.1187 & -1.2852$^{*}$ & -1.3741$^{*}$ & -1.4645$^{*}$\\ 
16 & -0.8377 & -0.8942 & -0.9567 & -1.0247 & -1.0977 & -1.2662$^{*}$ & -1.3556$^{*}$ & -1.4465$^{*}$\\  
18 & -0.8204 & -0.8765 & -0.9391 & -1.0073 & -1.0807 & -1.2505$^{*}$ & -1.3403$^{*}$ & -1.4315$^{*}$\\    
20 & -0.8063 & -0.8621 & -0.9246 & -0.9929 & -1.0666 & -1.2375$^{*}$ & -1.3275$^{*}$ & -1.4189$^{*}$\\  
\tableline
\end{tabular}
\caption{Ground state energy per site for the $4\times4$ cluster with particle density $n=0.75$. 
Stars indicate states with APBC and the representation of the point group 
that transforms like $s$-wave. Otherwise the states correspond to MBC.}
\label{table1}
\end{table}

\begin{table}
\begin{tabular}{ccccccc}
\tableline
\tableline
$U$ & $J=0$ & $J=0.25$ & $J=0.5$ & $J=0.75$ & $J=1$\\
0 & -1.6339 &  -1.6690 &  -1.7074 &  -1.7496 &  -1.7967\\  
2 & -1.2936 &  -1.3404 &  -1.3937 &  -1.4547 &  -1.5240\\ 
4 & -1.0473 &  -1.1091 &  -1.1812 &  -1.2625 &  -1.3514\\
6 & -0.8805 &  -0.9571 &  -1.0445 &  -1.1394 &  -1.2404$^{*}$\\ 
8 & -0.7712 &  -0.8580 &  -0.9540 &  -1.0566 &  -1.1685$^{*}$\\
10 & -0.6978 &  -0.7897 &  -0.8908 &  -0.9991$^{*}$ &  -1.1154$^{*}$\\
12 & -0.6460 &  -0.7406 &  -0.8446 &  -0.9568$^{*}$ &  -1.0751$^{*}$\\ 
14 & -0.6079 &  -0.7037 &  -0.8095 &  -0.9239$^{*}$ &  -1.0437$^{*}$\\
16 & -0.5788 &  -0.6752 &  -0.7819 &  -0.8979$^{*}$ &  -1.0186$^{*}$\\ 
18 & -0.5560 &  -0.6524 &  -0.7598 &  -0.8768$^{*}$ &  -0.9982$^{*}$\\ 
20 & -0.5376 &  -0.6338 &  -0.7417 &  -0.8594$^{*}$ &  -0.9812$^{*}$\\ 
\tableline
\end{tabular}
\caption{Ground state energy per site for the $4\times4$ cluster with particle density $n=0.875$.
Stars indicate states with APBC and the representation of the point group 
that transforms like $d_{x^2-y^2}$-wave. Otherwise the states correspond to MBC.} 
\label{table2}
\end{table}  
For fillings $n=0.75$ and $n=0.875$, the lowest energy in the
noninteracting case is achieved by considering MBC. 
In this limit,
 the only ingredient 
playing a role in the energetic  balance is the kinetic energy gain.
As interactions are 
switched on, the GS corresponds to APBC for sufficiently large $J$ and $U$.
This indicates that the ${\bf k}$-points tuned by  APBC are able to take advantage 
of some effect of the interactions,
compensating the loss of kinetic energy. For the latter boundary 
conditions,
the Fermi points of the noninteracting system lie on  
the lines of nodes of $f_{es}$.
Therefore, it is likely that the most favored instability by such a
 change of population in the ${\bf k}$-space
 is $d_{x^2-y^2}$-wave superconductivity.  
Some values of the GS energy per site are shown in tables \ref{table1} and \ref{table2}. 

It is interesting to note that, for the density $n=0.75$,
the GS belongs to the representation of the point symmetry group 
with $s$-wave-like character within the region where the GS
corresponds to APBC. Instead, for $n=0.875$ and also within the
region of parameters where the GS corresponds to APBC, the character of the
point-group representation  is $d_{x^2-y^2}$-wave like.
Since in the present cluster $n=0.875$, differs from $n=0.75$ in two  
particles, this change of representation is
consistent with the idea that a pair of particles with $d_{x^2-y^2}$-wave symmetry
was added to a many-particle background with total $s$-wave symmetry.
We actually speculate that such a background is also made up of paired
particles.

\begin{figure}
\includegraphics[width=65mm,angle=-90]{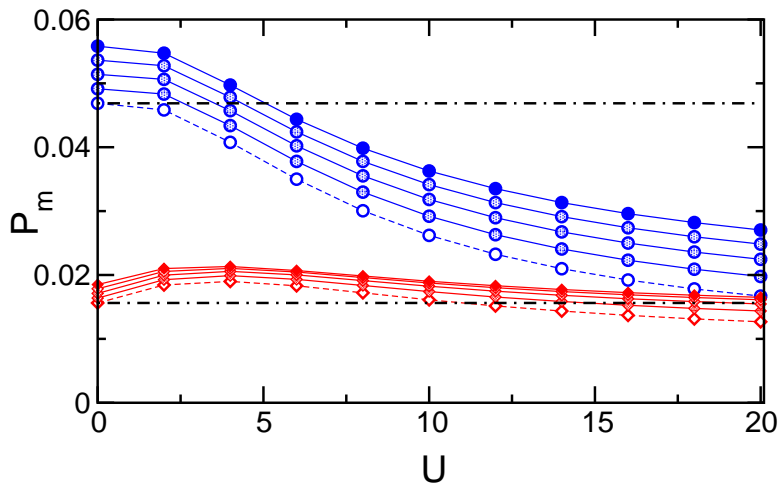}
  \caption{(Color online)
Pair correlation function at the maximum distance, $P_m$,
of the cluster for
a density of particles $n=0.625$. Blue circles correspond to pairs with $d_{x^2-y^2}$ 
and red diamonds to extended $s$ symmetry. Different plots correspond to $J=0,0.25,0.5,0.75,1$.
Open and filled dark symbols correspond to
the lowest and highest value of $J$, respectively.
 The dotted-dashed line indicates
the value of $P_m$ in the noninteracting limit. }
\label{p10}
\end{figure}
The behavior of $P_m$, the pair correlation function  (\ref{pcf}) corresponding to pairs 
separated by the maximum possible distance of the cluster, for 
a particle density 
$n=0.625$ is shown in Fig.\ref{p10}. The correlation function  corresponding to local pairs 
with $s$-wave symmetry is much weaker and is not shown. The corresponding values for the
noninteracting system are indicated in dot-dashed lines to ease the comparison. 
 In the latter limit, correlations of pairs with  $d_{x^2-y^2}$ symmetry
remain weaker than those of the noninteracting case while $s$-wave ones are slightly enhanced
for small enough $U$.
The effect of $J$ is to produce a weak enhancement of $P_m$ within
the two symmetry channels in comparison to the pure Hubbard
case. In particular, $d_{x^2-y^2}$ ones become  stronger than those of the 
noninteracting
case for small enough $U$.

\begin{figure}
\includegraphics[width=65mm]{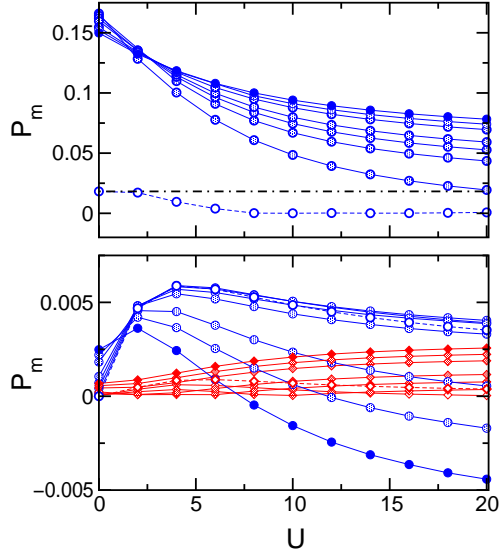}
  \caption{(Color online) Pair correlation function at the maximum distance
of the $4\times 4$ cluster, $P_m$, for a density of particles $n=0.75$. 
 Different plots correspond to $J=0,0.25,0.5,0.75,1,1.5,1.75,2$.
Upper and lower
panels correspond to APB and MBC respectively. Other details are as in Fig. \ref{p10}.}
\label{p12}
\end{figure}
The pairing function $P_m$ for the density $n=0.75$ is shown in Fig.\ref{p12}. We have 
analyzed the
behavior using the two boundary conditions leading to the lowest energy. In the case of 
APBC shown in
the upper panel of Fig. \ref{p12}, only the correlation of pairs with $d_{x^2-y^2}$ symmetry 
is shown,
since those with local s- and extended $s$-wave are negligibly small in comparison. 
 In the case of MBC shown in the lower panel, correlations in both, extended
$s$-wave and $d_{x^2-y^2}$-symmetry channels
 are only slightly enhanced for some values of $J$ and suppressed for others.
Instead, for APBC, a clear enhancement of
the correlations of pairs with $d_{x^2-y^2}$ is observed as $J$ is switched on. 
Note that, in contrast to the case $J=0$, the correlation function $P_m$ lies above
the line indicating the magnitude of $P_m$ in the noninteracting case.

\begin{figure}
 \epsfxsize=3.5in
  \epsffile{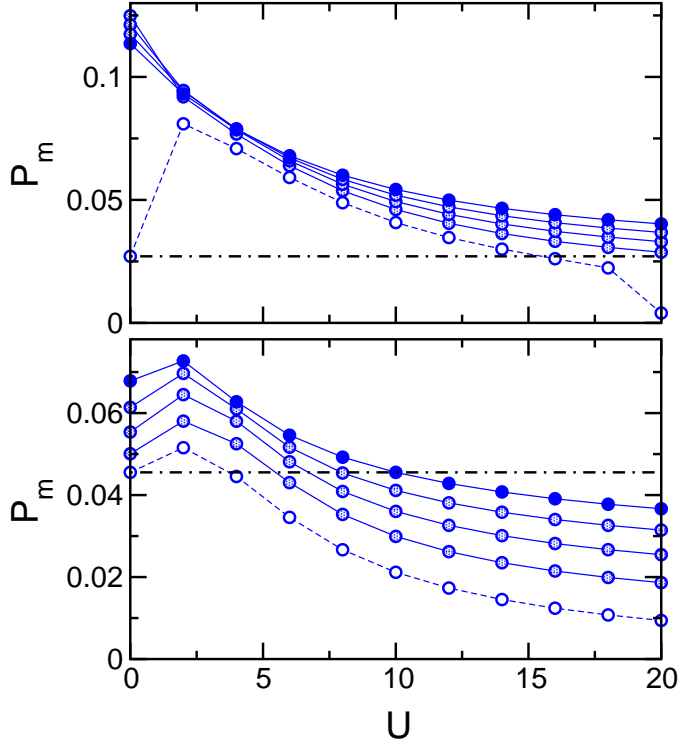}
  \caption{(Color online) Pair correlation function at the maximum distance
of the $4\times 4$ cluster, $P_m$, for a density of particles $n=0.875$. 
Other details are as in Fig. \ref{p12}.}
\label{p14}
\end{figure}
Similar remarks apply to the behavior of $P_m$ at the density $n=0.875$, shown in Fig.\ref{p14}.
The enhancement of $d_{x^2-y^2}$ pairing correlations for the case of APBC is 
related to the fact that the available ${\bf k}$ vectors mainly populated in the
ground state contribute with a sizable structure factor to a pairing interaction at the
Fermi surface with $d_{x^2-y^2}$ symmetry.

\begin{figure}
\includegraphics[width=65mm]{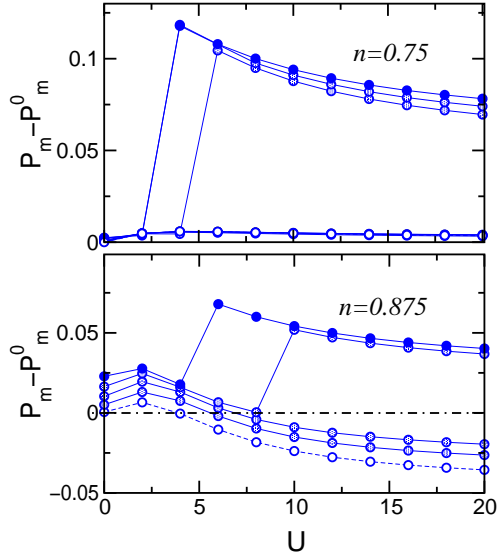}
  \caption{(Color online) Pair correlation function 
at the maximum distance
of the $4\times 4$ cluster, $P_m$, for pairs with $d_{x^2-y^2}$ symmetry
corresponding to the optimal boundary conditions for the ground state.
Upper and lower panels correspond to 
densities of particles $n=0.75$ and $n=0.875$, respectively. 
Other details are as in Figs. \ref{p10} and \ref{p12}.}
\label{pop}
\end{figure}
If, for the latter densities, we plot  the pair correlation function in the GS corresponding to
 the optimal boundary condition (i.e. that leading to the lowest energy), we obtain the
picture shown in Fig. \ref{pop}. 
In good agreement with the analysis done in the discussion about the behavior of the GS energy, 
we see that the change in the boundary condition leading to the lowest energy,
is accompanied with an enhancement of the $d_{x^2-y^2}$-wave pairing correlation
 function. The pairing correlation functions with extended $s$
symmetry are, instead, vanishing small within all the range of parameters. 
This could
be an unfortunate consequence of  
the small size of the cluster and to the fact that
the most populated ${\bf k}$-points when the boundary conditions change, lie on the lines of
nodes of the structure factor $f_{es}$. However, this behavior is in agreement with the results
predicted by RG. 

\begin{figure}
\includegraphics[width=65mm, angle=-90]{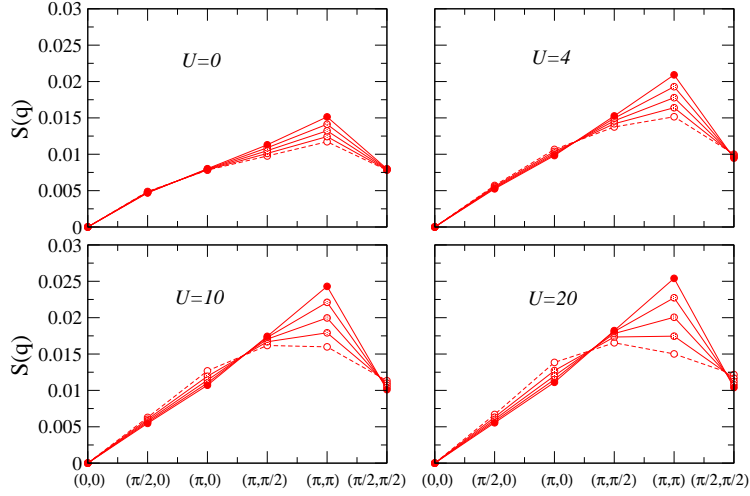}
  \caption{(Color online) Fourier transform of the spin-spin correlation function for
$n=0.75$ and MBC. 
Different panels correspond to different $U$. Open and filled dark
 symbols correspond to $J=0$ and $J=1$, respectively. Other plots 
correspond to intermediate equally spaced values of $J$.}
\label{sq075}
\end{figure}

\begin{figure}
\includegraphics[width=65mm,angle=-90]{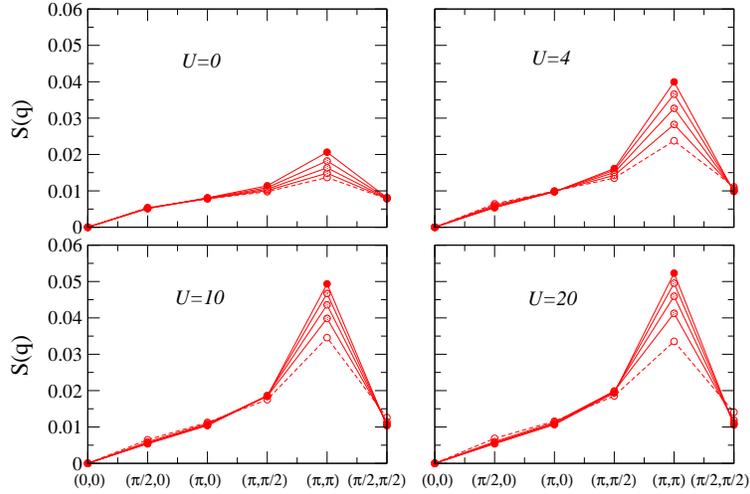}
  \caption{(Color online) The same as Fig. \ref{sq075} for $n=0.875$.}
\label{sq0875}
\end{figure}

To finalize, we present some results on the behavior of spin-spin correlation functions.
Fig. \ref{sq075} and Fig \ref{sq0875} show some typical plots  
corresponding to the GS in the cluster with MBC. In the case of Fig. \ref{sq075} the
latter boundary condition corresponds to the one leading to the lowest energy within the range of 
parameters shown. In  Fig. \ref{sq0875} this is not always the case  but we found that
there are only slight quantitative differences between the results of the figure and the
corresponding ones with APBC. The important feature to note is that the effect of $J$
is to increase the peak of  $S(\pi,\pi)$. By comparing the height of the latter
peak for the two densities it is clear that 
AF correlations increase as the system approaches to half-filling.
For $n=0.75$, $S({\bf k})$ shows a wide structure for the usual Hubbard model,
and for large $U$ the peak is placed at incommensurate positions  ${\bf k} \neq (\pi,\pi)$
(cf. lower panels of Fig. \ref{sq075}).  Remarkably, the effect of $J$ is to shift
these peaks to the AF vectors. All these features are consistent with the idea that
$J$ drives an enhancement of AF correlations relative to the usual Hubbard case.
The small size of the cluster does not allows us to have an estimate of the AF correlation
length. It can be, however, noted that for the lower density
(cf. Fig. \ref{sq075}) $S( {\bf k})$ spreads out on a wide range of ${\bf k}$ vectors
surrounding $(\pi,\pi)$, consistent with a picture of short-range AF correlations.
Instead, for lower doping, the structure evolves to a sharper peak around $(\pi,\pi)$
suggesting larger coherence lengths.
It is also interesting to note that the increment of AF correlations 
(from $n=0.75$ to $n=0.875$) is accompanied by a decrease of the pairing correlations (see Fig. 6),
in agreement with the RG results from the previous section.

\section{Summary and conclusions}
The main result of the present study is that the cooperative effect between
the nearest neighbor exchange interaction $J$ and the on-site Coulomb 
repulsion $U$
increases in a larger than simply additive way
the antiferromagnetic and $d_{x^2-y^2}$ superconducting tendencies in 2D.
For our analysis we used the angle resolved renormalization group
including self-energy corrections and the exact diagonalization methods.

We have first considered the repulsive and the attractive Hubbard models
and we have calculated the self-energy-dressed 
dominant correlation functions at half-filling
and at finite doping.
In the repulsive ($U > 0$) case self-energy effects
reduce radically all two-particle correlations and destroy their divergences
near the critical scale.
At half-filling the SC correlation function is below its $U=0$ value,
while the AF one remains stronger than its value for $U=0$, but loses the divergence.
This behavior, 
discussed in Ref.\cite{drazen2}, is a signature of the Mott localization tendencies with
simultaneous build-up of short ranged AF correlations.
At finite doping,
surprisingly and contrary to previous predictions made by the RG theory
without self-energy corrections (Ref.\cite{drazen1}), the superconducting 
instabilities are strongly reduced by self-energy corrections
even deeply in the BCS regime, i.e. when the Fermi surface is badly nested and umklapps
are irrelevant.
In the case of the attractive interaction ($U <  0$) our RG results are in complete agreement
with previous studies \cite{scal,ran}. At half-filling the $s$-wave superconductivity and the charge
density wave correlations are degenerate. The flow of the two correlation functions
looks similar to the one of the AF correlation function for repulsive Hubbard model at half-filling.
Just as in the repulsive case, the self-energy effect
regularizes the flow of correlation functions near $\Lambda_c$ at half-filling and no phase transition
occurs at this scale. The effective action of the regime below $\Lambda_c$ was discussed 
by Schulz \cite{Schulz_old}.
Contrary to the half-filled case, at finite doping  the attractive Hubbard model shows a convincing
on-set of the superconductivity. In comparison to the repulsive case where $d_{x^2-y^2}$ SC susceptibility
is only weakly enhanced, in the attractive case the $s$-wave
SC  correlations
at $\Lambda_c$ are not destroyed by the self-energy, while CDW susceptibility remains stuck to its 
$U=0$ value.
At this point some general remarks are in order. We have seen that there are fundamental differences
between RG flows of the repulsive and of the attractive Hubbard models.
The fluctuations in the repulsive model are much stronger and probably fatal for superconductivity.
They tend not only to decrease the magnitude of the SC susceptibility but also to 
decrease the energy $\Lambda^*=8t \exp(-l^*)$ at which it begins to depart from the behavior of
the noninteracting one.
Our present study
is unable to say if  the superconductivity is stabilized or not at some energy lower than
$\Lambda_c$.
However, the absence of all divergences indicates that the scenarios
with pre-formed pairs of AF and SC type are relevant even in the weak coupling limit.
These RG results agree with the ED analysis for the intermediate-to-strong couplings, where no
superconductivity was detected.
The situation is fundamentally different in the attractive Hubbard model, where the superconducting
instability is robust upon self-energy corrections. This is the RG version of the well known fact
that the incoherent preformed pairs can live only at intermediate-to-strong interaction
\cite{BCS-Bose}.

The study of the two Hubbard models was a necessary introduction to the RG analysis of the $t-J-U$
model. The question that we answer is if the superconductivity of the $t-J-U$ model
is enhanced or reduced with respect to the simple cases $J=0$ and $U=0$. The answer is that
$J$ and $U$ cooperate to
a) increase the critical cutoff
b) keeping the renornalized susceptibility enhanced relative to the
noninteracting one. We also found that the latter effect is observed not only on the
superconducting side of the phase diagram but also on the antiferromagnetic one. 
We thus have an evidence for the {\em U-J synergy effect}.
The phase diagram on Fig. \ref{pduj} shows in particular the cases ($U=1.6t, J=0$), ($U=0, J=1.6t$)
and ($U=1.6t, J=1.6t$). The critical temperature of the third case is much higher than for the
first two, while the onset of $d_{x^2-y^2}$ SC remains convincingly large in the flow of the correlation
function. 

Exact-diagonalization results support the above picture. In fact, 
for large enough $J$ and $U$, we find below half-filling a clear enhancement of the SC correlations
with $d_{x^2-y^2}$ symmetry which is accompanied by a change in the type of boundary conditions leading to
the lowest energy. This behavior is consistent with the idea that the large Coulomb repulsion
spreads out the Fermi surface towards sectors of the Brillouin zone where the interaction $J$ has
the largest amplitude in the BCS channel with $d_{x^2-y^2}$ symmetry.
 In such a way, we can think that particles are pushed 
to a region of the phase space, where the attractive interaction is most efficient to organize them
into pairs. Similar arguments can be proposed to explain the enhancement of AF correlations at 
half-filling since the $J$ interaction has components along AF and SC channels.   
In particular, in Ref. \cite{drazen2} it has been found that
the angle-resolved quasiparticle weight $Z(\theta)$ renormalizes in such a way that 
it displays a maximum in regions of the Fermi surface that are separated by the magnetic
vector ${\bf Q}=(\pi,\pi)$. We have not found strong differences between the behavior of
$Z(\theta)$ in the usual Hubbard model and the $t-J-U$ one neither at half-filling nor
for finite doping. Therefore, the most likely scenario at half-filling is that the 
 larger population of convenient regions of the ${\bf k}$-space  become available
to be exploited by the component of the  $J$ interaction along the AF channel.

We have not found any indication of bond-order-like instability. This is in disagreement with some
mean-field predictions \cite{laugh} but in full agreement with other RG studies on van-Hove fillings 
of the $t-t'-J-U$ models \cite{kat}. Our conclusion regarding the combined $J-U$ mechanism
to drive large superconducting correlations is also in agreement with previous investigations 
based on Fermi liquid arguments and quantum Monte Carlo simulations \cite{ple}.
Within the repulsive Hubbard model, we obtained by renormalization group
that the self-energy corrections are
fatal for superconductivity and by exact diagonalization that the
superconductivity is unlikely. This result is important because it reconciles
the N-patch RG with exact diagonalization {\em and} with a number of
other approaches. Our results show that while the $d_{x^2-y^2}$ pairing exists
in the Hubbard model (because vertex indeed diverges), the onset of
macroscopic
superconductivity is suppressed. This result is in agreement with
the very recent findings by Plekhanov and co-workers \cite{Plekhanov_2004}.

All above remarks indicate that the $J$ interaction is vital for the superconductivity while the
$U$ interaction increases  considerably the tendency towards pair formation. 
They also suggest that minimal microscopic models supporting recent phenomenological 
proposals for colossal effects in the phase diagram of the high $T_c$ compounds \cite{col}
 may be based on these two interactions.
In fact, our results indicate that the interaction $J$ provides a kick to the
potential or weak antiferromagnetic and superconducting tendencies of the Hubbard model,
that triggers a gigantic response in the system. 
This picture resembles the behavior of manganites
where colossal magnetoresistance effects take place in response to external magnetic fields and
this analogy is behind the proposal of Ref. \cite{col}. The appropriate microscopic approach
should be based on a recently developed N-patch renormalization group theory
for disordered and interacting imperfectly nested system \cite{dusuel04}.
On the other hand, as discussed in our introductory section, derivations starting from the 
three band model for the cuprates also support the idea that the
$t-J-U$ model is a good candidate for  a minimal one-band Hamiltonian for these materials.

\section{Acknowledgements} 
We thank Elbio Dagotto for encouraging remarks.   
LA thanks Prof. Fulde for his kind hospitality at the Max-Planck Institut
f\"ur Physik komplexer Systeme, Dresden, where a good part of this work has been
carried out, as well as the support 
of the Alexander von Humboldt Stiftung and CONICET, Argentina. 
Support from PICT 03-11609 is also acknowledged.


\begin{thebibliography}{999}
\bibitem{note}The effective model is actually more general, containing in addition
a nearest-neighbor Coulomb interaction $V$, a hopping of local pairs $X$ and a 
bond-charge interaction $\Delta$. However, the only interaction driving superconductivity
with $d_{x^2-y^2}$ symmetry at mean-field level is $J$ and for this reason this is the only one we keep 
in  the present work.
\end{thebibliography}
\end{document}